\documentclass[aps,prb,twocolumn,superscriptaddress,showpacs,amsmath,amssymb,
footinbib,longbibliography,10pt]{revtex4-2}
\usepackage[utf8]{inputenc}
\usepackage{bm}
\usepackage[usenames]{color}
\usepackage{multirow}
\usepackage{amssymb}
\usepackage{amsbsy}
\usepackage{amsmath}
\usepackage{stmaryrd}
\usepackage{graphicx}
\usepackage{epsfig}
\usepackage{placeins}
\usepackage[normalem]{ulem}
\usepackage{bbold}
\usepackage{bbm}
\usepackage{braket}
\usepackage{graphicx}
\usepackage{tikz}
\usepackage{color}

\usetikzlibrary{calc}
\usetikzlibrary{patterns}
\usepackage[colorlinks,linkcolor=blue,citecolor=blue,urlcolor=blue]{hyperref}

\makeatletter

\setcitestyle{numbers,square,comma,sort&compress}

\nocite{*}

\definecolor{palette1}{HTML}{A8216B}
\definecolor{palette2}{HTML}{F1184C}
\definecolor{palette3}{HTML}{F36943}
\definecolor{palette4}{HTML}{F7DC66}
\definecolor{palette5}{HTML}{2E9599}

\usepackage{scalerel}
\usepackage{tikz}
\usetikzlibrary{calc}
\usetikzlibrary{patterns}
\usetikzlibrary{svg.path}
\definecolor{orcidlogocol}{HTML}{A6CE39}
\tikzset{
  orcidlogo/.pic={
    \fill[orcidlogocol]
svg{M256,128c0,70.7-57.3,128-128,128C57.3,256,0,198.7,0,128C0,57.3,57.3,0,128,
0C198.7,0,256,57.3,256,128z};
    \fill[white] svg{M86.3,186.2H70.9V79.1h15.4v48.4V186.2z}

svg{M108.9,79.1h41.6c39.6,0,57,28.3,57,53.6c0,27.5-21.5,53.6-56.8,53.6h-41.8V79.
1z
M124.3,172.4h24.5c34.9,0,42.9-26.5,42.9-39.7c0-21.5-13.7-39.7-43.7-39.7h-23.
7V172.4z}

svg{M88.7,56.8c0,5.5-4.5,10.1-10.1,10.1c-5.6,0-10.1-4.6-10.1-10.1c0-5.6,4.5-10.1
,10.1-10.1C84.2,46.7,88.7,51.3,88.7,56.8z};
  }
}

\newcommand\orcid[1]{\!%
  \href{https://orcid.org/#1}{%
    \mbox{%
      \scaleto{%
        \begin{tikzpicture}[yscale=-1,transform shape]
          \pic{orcidlogo};
        \end{tikzpicture}
      }{8pt}%
    }%
  }%
}

\begin{document}

\title{Scaling of diffusion constants in perturbed easy-axis Heisenberg spin
chains}

\author{Markus Kraft~\orcid{0009-0008-4711-5549}}
\email{markus.kraft@uos.de}
\affiliation{University of Osnabr{\"u}ck, Department of Mathematics/Computer
Science/Physics, D-49076 Osnabr{\"u}ck, Germany}

\author{Mariel Kempa~\orcid{0009-0006-0862-4223}}
\affiliation{University of Osnabr{\"u}ck, Department of Mathematics/Computer
Science/Physics, D-49076 Osnabr{\"u}ck, Germany}

\author{Jiaozi Wang~\orcid{0000-0001-6308-1950}}
\affiliation{University of Osnabr{\"u}ck, Department of Mathematics/Computer
Science/Physics, D-49076 Osnabr{\"u}ck, Germany}

\author{Sourav Nandy~\orcid{0000-0002-0407-3157}}
\affiliation{Max Planck Institute for the Physics of Complex Systems, D-01187
Dresden, Germany}

\author{Robin Steinigeweg~\orcid{0000-0003-0608-0884}}
\email{rsteinig@uos.de}
\affiliation{University of Osnabr{\"u}ck, Department of Mathematics/Computer
Science/Physics, D-49076 Osnabr{\"u}ck, Germany}

\date{\today}


\begin{abstract}
Understanding the physics of the integrable spin-$1/2$ XXZ chain has witnessed
substantial progress, due to the development and application of sophisticated
analytical and numerical techniques. In particular, infinite-temperature
magnetization transport has turned out to range from ballistic, over
superdiffusive, to diffusive behavior in different parameter regimes of the
anisotropy. Since integrability is rather the exception than the rule, a crucial
question is the change of transport under integrability-breaking perturbations.
This question includes the stability of superdiffusion at the isotropic point
and the change of diffusion constants in the easy-axis regime. In our work, we
study this change of diffusion constants by a variety of methods and cover both,
linear response theory in the closed system and the Lindblad equation in the
open system, where we throughout focus on periodic boundary conditions. In the
closed system, we compare results from the recursion method to calculations for
finite systems and find evidence for a continuous change of diffusion constants
over the full range of perturbation strengths. In the open system weakly
coupled to baths, we find diffusion constants in quantitative agreement with
the ones in the closed system in a range of nonweak perturbations, but
disagreement in the limit of weak perturbations. Using a simple model in this
limit, we point out the possibility of a diverging diffusion constant in such
an open system.
\end{abstract}

\maketitle


\section{Introduction} The study of nonequilibrium processes in quantum
many-body
systems continues to be one of the central endeavors in different fields
of modern physics \cite{Bloch2008, Polkovnikov2011, Eisert2015, DAlessio2016,
Abanin2019}, ranging from fundamental questions in statistical mechanics to
applied questions in material science. In this context, transport is a
paradigmatic example of a nonequilibrium process with relevance to closed and
open quantum systems alike. The understanding of nonequilibrium physics
in general and transport in particular has seen substantial
progress \cite{Bertini2021}, due to experimental
advances, fresh theoretical concepts, and the development of sophisticated
analytical techniques and numerical methods.

Within the large class of physically relevant quantum many-body models,
integrable systems are rare and play a special role. In particular, the
spin-1/2 XXZ chain is a prime example of an integrable system and has been
scrutinized in numerous studies. Early on, it has become clear that this model
can fail to thermalize, which is also reflected in the ballistic flow of energy
\cite{Zotos1997}. Other observables than energy,
however, can have a richer dynamical phase diagram, even at high temperatures.
Especially the flow of magnetization is ballistic in the easy-plane regime
only, where quasilocal conserved charges have an overlap with the current
\cite{Prosen2011, Prosen2013, Ilievski2016}. Otherwise,
the transport of magnetization has turned out to be superdiffusive at the
isotropic point \cite{Znidaric2011, Ilievski2018,Ljubotina2019} and diffusive in
the easy-axis regime
\cite{Prelovsek2004,Michel2008,Prosen2009,Steinigeweg2011,Karrasch2014,
DeNardis2019, Gopalakrishnan2019}.

Since integrable systems are rather the exception than the rule, nonintegrable
systems are the generic situation in physics. While these systems are expected
to typically exhibit diffusive transport, the precise value of diffusion
constants constitutes a highly nontrivial question and is a challenge for
theory \cite{Steinigeweg2014, Kloss2018,
Richter2019, Wurtz2020, Rakovszky2022, Wang2023, Yi-Thomas2024,
Artiaco2024}. An equally nontrivial question is the dynamical behavior in
the close vicinity of integrability \cite{Jung2006,
Jung2007, Steinigeweg2016, Brenes2018, Znidaric2020, Bastianello2021,
Znidaric2022}. This question includes the stability of
superdiffusion at the isotropic point \cite{DeNardis2023,
Roy2023, Nandy2023, Gopalakrishnan2024, McRoberts2024}, but also the change
of diffusion constants in the easy-axis regime
\cite{DeNardis2022, Prelovsek2022,Pawlowski2025}, where this change
might be discontinuous. While such a change is in line with breaking
microscopic integrability, it is still unexpected from macroscopic
phenomenology, where a small perturbation of a normal conductor does not cause
a significant change of the transport coefficient.

In our work, we study this change of diffusion constants by a variety of
methods and cover both, linear response theory in the closed system and the
Lindblad equation in the open system, where we throughout focus on periodic
boundary conditions. In the closed system, we compare results from the
recursion method \cite{Wang2023} to exact results for
finite systems and find evidence for a continuous
change of diffusion coefficients over the full range of perturbation strengths.
In the open system, we find diffusion constants in quantitative agreement with
the ones in the closed system for nonweak perturbations, but disagreement for
weak perturbations. Using a simple model, we point out that the open-system
diffusion constant might diverge in the weak-perturbation limit.


\section{Model and methods} In our work, a central model is the
spin-$1/2$ XXZ chain \cite{Bertini2021},
\begin{equation} \label{eq:model}
H = J \sum_{r=1}^{N} \left ( S_{r}^{x} S_{r+1}^{x} +
S_{r}^{y} S_{r+1}^{y} +\Delta S_{r}^{z}S_{r+1}^{z} \right ) \, ,
\end{equation}
where the $S_{r}^{j}$ $(j=x,y,z)$ are spin-$1/2$ operators at site $r$, $N$
is the number of sites, $J > 0$ is the antiferromagnetic exchange coupling
constant, and $\Delta$ is the anisotropy in the $z$ direction. For all values of
$\Delta$, the model in Eq.\ (\ref{eq:model}) is integrable and the total
magnetization $S^z = \sum_r S_r^z$ is conserved, $[S^z, H]=0$. Throughout
our work, we focus on the easy-axis regime $\Delta > 1$ and choose the specific
value $\Delta = 1.5$. It is important to note that we employ periodic boundary
conditions, $S_{N+1}^{j} = S_{1}^{j}$, in the closed system but also in the open
system introduced later.

To break the integrability of the model, we consider two different
perturbations. First, interactions between next-to-nearest sites,
\begin{equation} \label{eq:model_NNN}
H_{\text{NNN}} = H +\Delta^{\prime} \sum_{r=1}^{N} S_{r}^{z}S_{r+2}^{z} \, ,
\end{equation}
and, second, a staggered magnetic field,
\begin{equation} \label{eq:model_B}
H_{\text{B}} = H + B \sum_{r=1}^{N} (-1)^{r} S_{r}^{z} \, .
\end{equation}
Here, the parameters $\Delta^{\prime}$ and $B$ are the corresponding
perturbation strengths. For all values of $\Delta^{\prime}$ and $B$, also
including the case $\Delta^{\prime} = B = 0$, transport is expected to be
diffusive in the easy-axis regime.

In the closed system, we rely on linear response theory (LRT) \cite{Kubo1991},
where an essential role is played by the operator of the magnetization current
\cite{Bertini2021}
\begin{equation}
j = J \sum_{r=1}^{N} \left ( S_{r}^{x} S_{r+1}^{y} -
S_{r}^{y} S_{r+1}^{x} \right )
\end{equation}
and its autocorrelation function
\begin{eqnarray} \label{eq:correlation_function}
\langle j j(t)\rangle  = \frac{\text{tr}[e^{-\beta H} e^{iHt} j e^{-iHt}
j]}{Z} \, , \quad Z = \text{tr}[e^{-\beta H}] \, ,
\end{eqnarray}
where $\beta = 1/T$ is the inverse temperature. In our work, we focus on the
high-temperature case $\beta = 0$. To study diffusion constants, we define the
quantity \cite{Steinigeweg2009b}
\begin{eqnarray} \label{eq:D_LRT}
D(t) = \frac{1}{\chi} \int_{0}^{t} \text{d}t' \, \langle j(t') j \rangle \, ,
\end{eqnarray}
with the static susceptibility $\chi =1 /4$ for $\beta = 0$. In this way,
diffusion constants $D = \lim_{t \to \infty} \lim_{N \to \infty} D(t)$ can be
obtained, where the order of limits is crucial. In practice, however, one has
often access to finite systems only, and then a finite-size scaling is
unavoidable. In such a finite-size scaling, the choice of a proper time scale
can be a subtle task, as discussed later in detail. Moreover, this finite-size
scaling can also depend on the choice of the specific ensemble, and
extrapolations can differ for the grandcanonical ensemble $\langle S^z \rangle
= 0$ (all magnetization sectors) and the canonical ensemble $S^z = 0$ (only
zero magnetization sector).

In our work, we apply dynamical quantum typicality for the numerical calculation
of the quantity in Eq.\ (\ref{eq:D_LRT}), which allows us to treat
comparatively large system sizes outside the range of standard exact
diagonalization \cite{Heitmann2020, Jin2021}. This
approach to finite systems is complemented by another technique, which is
designed to numerically calculate $D$ directly in the thermodynamical limit.
This technique is a recently proposed recursion method (RM) \cite{Wang2023} and
is based on Lanczos coefficients in Liouville space. Details of this method can
be found in the Appendix \ref{ap::A}.

\begin{figure}[t]
\includegraphics[width=0.8\columnwidth]{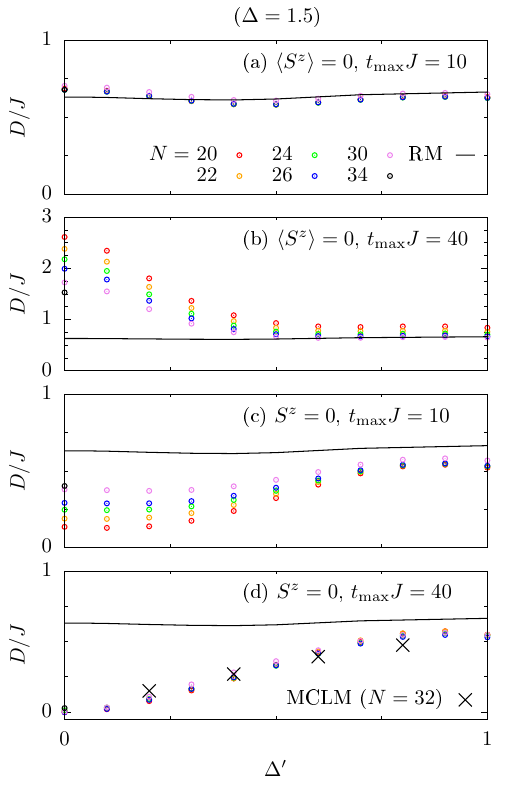}
\caption{Diffusion constant $D$ versus perturbation strength $\Delta^{\prime}$
for the model $H_{\text{NNN}}$ in Eq.\ (\ref{eq:model_NNN}) with anisotropy
$\Delta = 1.5$, as calculated numerically by dynamical quantum typicality for
different system sizes $N$ and by the recursion method in the thermodynamic
limit \cite{line}. (a) $\langle S^{z}\rangle = 0$,
$tJ_{\text{max}}=10$. (b) $\langle S^{z}\rangle = 0$, $tJ_{\text{max}}=40$. (c)
$S^{z} = 0$, $tJ_{\text{max}}=10$. (d)
$S^{z} = 0$, $tJ_{\text{max}}=40$. In (d), existing results from the
microcanonical Lanczos
method are also depicted
\cite{Prelovsek2022}.}
\label{fig:diffusion_NNN}
\end{figure}

\begin{figure}[b]
\includegraphics[width=0.8\columnwidth]{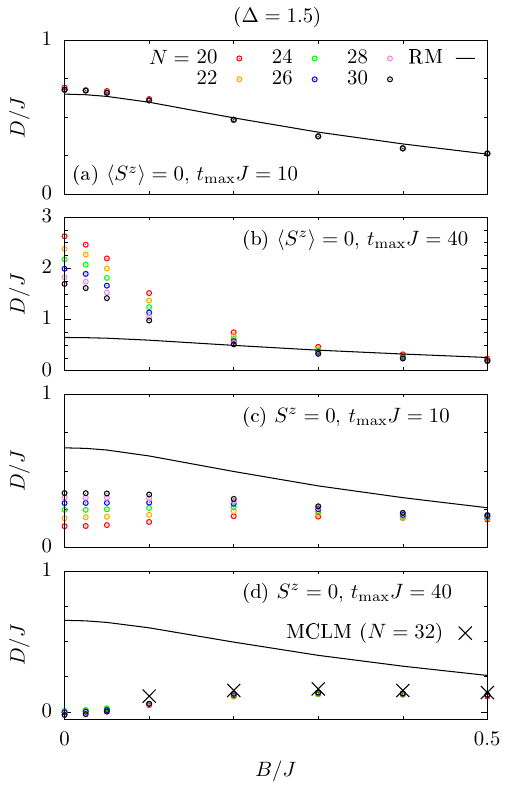}
\caption{Similar data as the one in Fig.\ \ref{fig:diffusion_NNN}, but now for
the model $H_{\text{B}}$ in Eq.\ (\ref{eq:model_B}).}
\label{fig:diffusion_staggered}
\end{figure}

In addition to the closed systems discussed so far, we further consider
open systems and study nonequilibrium transport, as resulting by
the coupling to two baths. To this end, we use the Lindblad equation
\cite{Breuer2007}
\begin{equation} \label{eq:Lindblad}
\dot{\rho}(t) = \mathcal{L}\rho(t) = i [\rho(t),H] + \mathcal{D} \rho(t) \, ,
\end{equation}
where the first term on the r.h.s.\ describes the unitary time evolution of a
density matrix $\rho(t)$ in the isolated situation, while the second term on the
r.h.s.\ describes nonunitary damping due to the baths and reads
\begin{equation}
\mathcal{D} \rho(t) = \sum_{j} \alpha_{j} \left(
L_{j}\rho(t)L_{j}^{\dagger} - \frac{1}{2} \left \{  \rho(t),L_{j}^{\dagger}L_{j}
\right \} \right)
\end{equation}
with non-negative rates $\alpha_{j}$, Lindblad operators $L_{j}$, and the
anticommutator $\{\bullet,\bullet \}$. The Lindblad equation is the most general
form of a quantum master equation, which is local in time and maps a density
matrix to a density matrix again \cite{Breuer2007}. In the specific context of
nonequilibrium
transport, a common choice for the Lindblad operators is given by
\cite{Bertini2021}
\begin{equation}
L_{1} = S^{+}_{1} \, , \, L_{2}= S^{-}_{1} \, , \, L_{3} = S^{+}_{N/2 +1} \, ,
\, L_{4} = S^{-}_{N/2 +1} \, ,
\end{equation}
which, in our case of periodic boundary conditions, are located at site $r = 1$
and $r = N/2 +1$. The corresponding rates read $\alpha_{1} = \gamma(1+\mu),
\alpha_{2} = \gamma(1-\mu)$, $\alpha_{3} = \gamma(1-\mu)$,
$\alpha_{4} = \gamma(1+\mu)$, where $\gamma$ is the system-bath coupling and $
\mu$ is the driving strength. In this setup, a nonequilibrium steady state
results in the long-time limit, which features a constant current and a
characteristic density profile. In case of normal transport, the diffusion
constant is given by the ratio
\begin{equation}
D = - \frac{\langle j_r \rangle}
{\langle S_{r+1}^{z}\rangle - \langle S_{r}^{z}\rangle}
\end{equation}
for some site $r$ in the bulk. We determine
this diffusion constant for small-system bath coupling $\gamma/J = 0.1$ and
weak driving $\mu = 0.1$.

\begin{figure}[t]
\includegraphics[width=0.8\columnwidth]{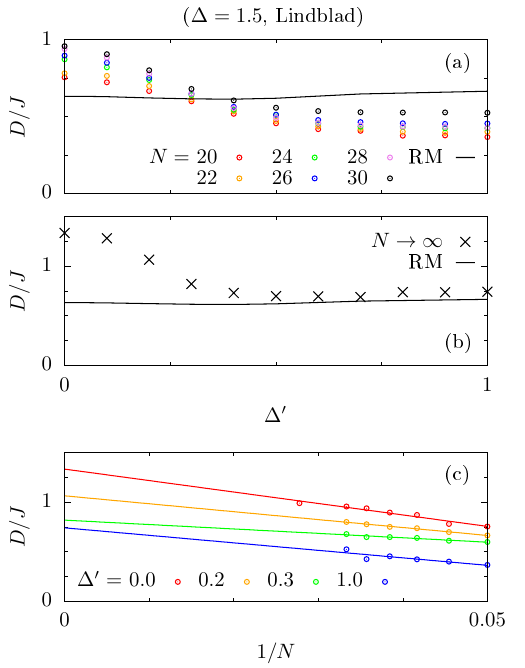}
\caption{(a) Diffusion constant $D$ versus perturbation strength
$\Delta^{\prime}$ for the model $H_{\text{NNN}}$ in Eq.\ (\ref{eq:model_NNN})
with anisotropy $\Delta = 1.5$, as calculated numerically in the open system
for different system sizes $N$. A comparison to the closed-system results from
the recursion method is also indicated. (b) Extrapolation to the
thermodynamic limit based on the linear scaling in (c).}
\label{fig:diffusion_open}
\end{figure}

A practical advantage of the Lindblad equation is that it can be solved by
sophisticated numerical techniques, like stochastic unraveling
\cite{Dalibard1992, Michel2008} or simulations based
on matrix product states (MPS) \cite{Vidal2003, Vidal2004, Verstraete2008}. In
particular, MPS simulations are feasible for quite large system sizes. Yet,
they are restricted to open boundary conditions and to large values of the
system-bath coupling $\gamma$, due to the required reduction of entanglement
growth. Thus, we use as an alternative strategy a recently proposed approach
\cite{Heitmann2023a, Heitmann2023, Kraft2024}. On the one hand, this
approach is suitable for periodic boundary conditions and small system-bath
coupling $\gamma$ \cite{gamma}. On the other hand, it allows us to predict the
open-system
dynamics just on the knowledge of spatio-temporal correlation functions in the
closed systems, which in turn enables us to treat open systems outside the
range of exact diagonalization and stochastic unraveling. Details on this
approach can be found in the Appendix \ref{ap::B}.


\section{Results} We now move on to our results, starting with the
closed-system
scenario and the model in Eq. (\ref{eq:model_NNN}). We first focus on
results from the recursion method, which are, as discussed in the Appendix
\ref{ap::A},
well converged w.r.t.\ to the number of Lanczos coefficients and
depicted in Fig.\ \ref{fig:diffusion_NNN} (a). As clearly visible, the
diffusion constant $D$ has a continuous dependence on
the perturbation strength $\Delta'$, with no obvious signature of a
discontinuity in the weak-perturbation limit $\Delta' \to 0$. Furthermore, for
the values of $\Delta'$ considered, $D(\Delta')$ has a minor dependence on
$\Delta'$ and is an almost flat curve.

In addition, we compare in Fig.\ \ref{fig:diffusion_NNN} (a)
the results from the recursion method to finite-size results for the
quantity in Eq.\ (\ref{eq:D_LRT}) from dynamical quantum typicality, for which
raw data can be found in the Appendix \ref{ap::A}. We do so for the
grandcanonical ensemble $\langle S^z \rangle = 0$ and at a time $t J = 10$.
Apparently, the finite-size data is in very good agreement with the recursion
method and shows no significant dependence on system size. When redoing the
comparison for a longer time $t J = 40$ in Fig.\ \ref{fig:diffusion_NNN} (b),
the agreement becomes worse, due to finite-size effects at nonweak perturbation
strengths $\Delta'$. Still, when the system size is increased, the
finite-size data gets closer and closer to the result from the recursion
method, and they might eventually coincide in the thermodynamic limit.

\begin{figure}[b]
\includegraphics[width=0.8\columnwidth]{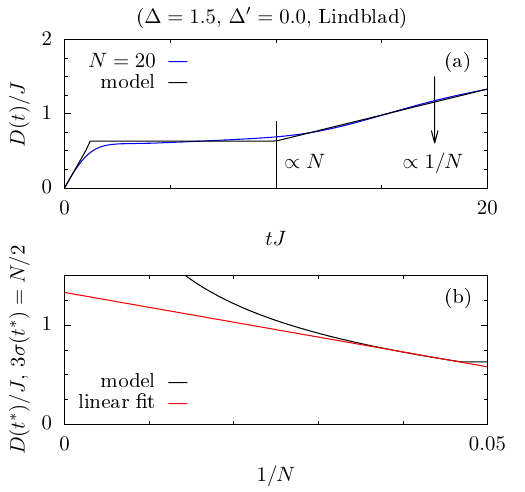}
\caption{(a) A simple model for the quantity $D(t)$ in Eq.\ (\ref{eq:D_LRT}):
After an initial linear increase, $D(t)$ has a plateau but, after a finite-size
time $\propto N$, increases linearly with a finite-size slope $\propto 1/N$.
Free constants are fixed by a fit to $N = 20$ data for the unperturbed case
in Eq.\ (\ref{eq:model}). (b) Resulting prediction for the finite-size scaling
of the diffusion constant $D$ in the open systems.}
\label{fig:model}
\end{figure}

It is instructive to redo the same comparison for the canonical ensemble $S^z =
0$ as well. As shown in Fig.\ \ref{fig:diffusion_NNN} (c) for a time $t J
= 10$, finite-size effects are significantly larger than before,
but they seem to be still consistent with the recursion
method. Interestingly, for a longer time $t J = 40$ in Fig.\
\ref{fig:diffusion_NNN} (d), finite-size effects seem to be smaller. However,
this observation has to taken with special care, in view of the finite-size
effects at shorter times.

In Figs.\ \ref{fig:diffusion_staggered} (a)-(d), we repeat the analysis for the
other model in Eq.\ (\ref{eq:model_B}). While $D(B)$ has a stronger dependence
on $B$, the overall behavior is very similar to the one of $D(\Delta')$ in
Figs.\ \ref{fig:diffusion_NNN} (a)-(d), which indicates that our results do not
depend on the specific perturbation.

Next, we turn to the open-system scenario and focus on the model in Eq.
(\ref{eq:model_NNN}) again. In Fig.\ \ref{fig:diffusion_open} (a), we summarize
our results for the diffusion constant $D(\Delta')$, which we obtain
from current and density profile of the steady state in the long-time limit.
Apparently, $D(\Delta')$ differs from the result of the recursion method and
increases with system size. To conclude on the thermodynamic limit, we show
in Fig.\ \ref{fig:diffusion_open} (b) an extrapolation, as resulting from linear
fits in Fig.\ \ref{fig:diffusion_open} (c). While this extrapolation for
$D(\Delta')$ agrees convincingly with the result of the recursion method in a
range of nonweak perturbation strengths $\Delta'$, it clearly differs in the
limit of weak perturbations $\Delta' \to 0$. This observation suggests that
open and closed systems have different diffusion constants in this limit.

\section{Simple model} To understand this observation, let us for a moment come
back to the closed system and use a simple model for the quantity $D(t)$ in Eq.\
(\ref{eq:D_LRT}), as shown in Fig.\ \ref{fig:model}. After an initial linear
increase, $D(t)$ has a plateau but, after a finite-size time $\propto N$,
increases linearly with a finite-size slope $\propto 1/N$. This overall
behavior describes well the unperturbed case $\Delta' = 0$ with periodic
boundary conditions \cite{Steinigeweg2012}. It is
worth mentioning that the linear slope is the finite-size Drude weight
\cite{Bertini2021}. Using this model for $D(t)$, we
can calculate the spreading of an excitation,
\begin{equation} \label{eq:variance}
\sigma^2(t) = 2 \int_0^t \text{d}t' \, D(t') \, ,
\end{equation}
and determine the time $t^*$ required to travel through the system,
\begin{equation}
3 \sigma(t^*) = \frac{N}{2} \, .
\end{equation}
Assuming that $t^*$ is the relevant time scale for the open system, we plug
in $t^*$ in the original model for $D(t)$ and use $D(t^*)$ as an prediction for
the diffusion constant in the open system. In Fig.\ \ref{fig:model} (b),
we depict the finite-size scaling of the so obtained $D(t^*)$, which is,
despite the simplicity of arguments, very similar to the one for the
open-system diffusion constant in Fig.\ \ref{fig:diffusion_open} (c).
Importantly, however, $D(t^*)$ departs from a $1/N$-scaling for large $N$
and eventually diverges in the limit $N \to \infty$. The
departure results from the fact that the steady-state time $t^*$ increases
faster with $N$ than the time at which $D(t)$ is converged w.r.t.\ system
size, see Appendix \ref{ap::C}. This finding opens
the possibility of a diverging diffusion constant
in the unperturbed open system with periodic boundary
conditions. Additional results for open boundary conditions can be found in
the Appendix \ref{ap::B}.


\section{Conclusion} We have analyzed the scaling of diffusion constants under
integrability-breaking perturbations of the spin-$1/2$ XXZ chain in the
easy-axis regime. To this end, we have used a variety of methods and covered
both, closed and open systems, focusing on periodic boundary conditions. In the
closed system, we have found evidence for a continuous change of diffusion
constants over the full range of perturbation strengths. In the open system, we
have found diffusion constants in quantitative agreement with the ones in the
closed system in a range of nonweak perturbations. In the weak-perturbation
limit, however, we have found disagreement and also pointed out the possibility
of a diverging diffusion constant in the open system.


\section*{Acknowledgments} We thank Jochen Gemmer, Jacek
Herbrych, Zala Lenar\v{c}i\v{c}, Marcin Mierzejewski, and Peter Prelov\v{s}ek
for fruitful discussions.

This work has been funded by the Deutsche
Forschungsgemeinschaft (DFG), under Grant No.\ 531128043, as well as under Grant
No.\ 397107022, No.\ 397067869, and No.\ 397082825 within DFG Research Unit FOR
2692, under Grant No.\ 355031190.

Additionally, we greatly acknowledge computing time on the
HPC3 at the University of Osnabrück, granted by the DFG, under Grant
No.\ 456666331. Computations for open boundary conditions were performed at the
HPC cluster facility at MPI-PKS Dresden.

\appendix



\section{Calculation of equilibrium transport}\label{ap::A}


\begin{figure}[t]
\includegraphics[width=0.9\columnwidth]{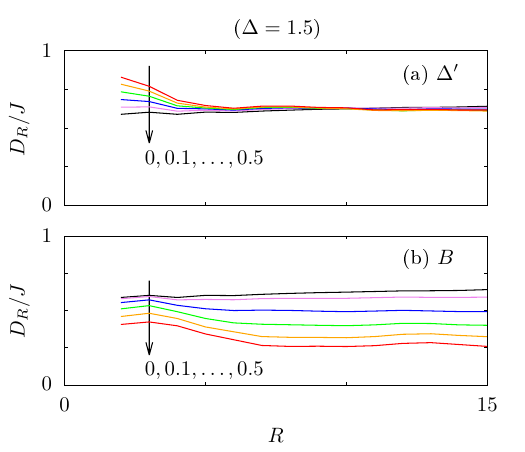}
\caption{Diffusion constant $D_R$ versus $R$, as obtained from the recursion
method for the model (a) $H_\text{NNN}$ in Eq.\ (\ref{eq:model_NNN}) and (b)
$H_\text{B}$ in Eq.\ (\ref{eq:model_B}).}
\label{fig::DR}
\end{figure}

\subsection{Recursion method}

Here, we briefly introduce the recursion method, which is used in the main text
for the calculation of diffusion coefficients. It is aiming at a calculation
directly in the thermodynamical limit. More details on the method
can be found in Ref.\ \cite{Wang2023}.

Within this framework, the Laplace transform
\begin{equation} \label{eq-Fs}
{F}(s) = \int_{0}^{\infty} \text{d}t \, e^{ts} \, \tilde{C}(t)
\end{equation}
is considered for the autocorrelation function
\begin{equation}
\tilde{C}(t) = \langle \tilde{j}(t) \tilde{j} \rangle \, , \quad
\tilde{j} = \frac{j}{\sqrt{\langle
j|j\rangle}}
\end{equation}
of the normalized current. In particular, ${\cal F}(0)$ determines the
diffusion constant $D=\langle j^{2}\rangle F(0)/\chi$. According to the Mori
formalism, Eq.\ \eqref{eq-Fs} can be written in the form
\begin{equation}
{F}(s) = \frac{1}{s + \frac{b_{1}^{2}}{s + \frac{b_{2}^{2}}{s +
\frac{b_{3}^{2}}{\cdots}}}} \, ,
\end{equation}
and one straightforwardly gets
\begin{equation}
D = \frac{\langle j^{2} \rangle}{\chi b_{1}} \prod_{n=1}^{\infty} \left(
\frac{b_{n}}{b_{n+1}} \right)^{(-1)^{n}} \, ,
\end{equation}
where $b_n$ are the Lanczos coefficients of $\langle
\tilde{j} \rangle$, the meaning of which will be explained below.

It is now convenient to switch to the Hilbert space of operators and denote its
elements $O$ by states $|O)$. This space is equipped with an inner product,
\begin{equation}
(O_m|O_n) = \text{tr}[O^\dagger_m
O_n] \, ,
\end{equation}
which defines a norm via
\begin{equation}
|| O || = \frac{\sqrt{(O|O)}}{\text{tr}[1]} \, .
\end{equation}
The Liouvillian (super)operator ${\cal L}$, which is defined
by ${\cal L} \, |O) =
[H, O]$,
propagates a state $|O)$ in time, such that an autocorrelation function can be
written as
\begin{equation}
\langle O(t) O \rangle \propto C(t) = \frac{(O|e^{i{\cal L}t}|O)}{|| O
||^2} \, .
\end{equation}

\begin{figure}[t]
\includegraphics[width=0.8\columnwidth]{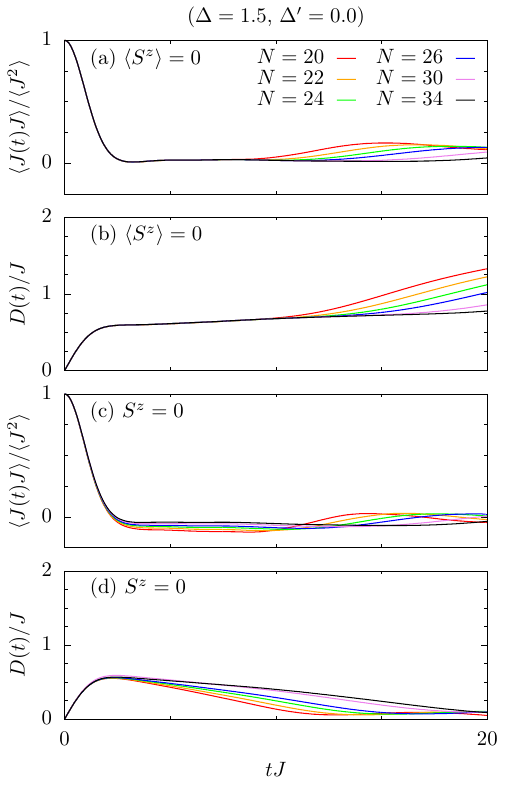}
\caption{Decay of the current autocorrelation function $\langle j(t) j
\rangle$ in the unperturbed model in Eq.\ (\ref{eq:model}) with anisotropy
$\Delta = 1.5$ for (a) the grandcanonical ensemble $\langle S^z \rangle = 0$
and (c) the canonical ensemble $S^z = 0$, which is numerically calculated for
different system sizes $N$ by the use of dynamical quantum typicality. (b) and
(d) Corresponding time dependence of the diffusion coefficients $D(t)$.
} \label{fig::finit_size_H}
\end{figure}

\begin{figure}[t]
\includegraphics[width=0.8\columnwidth]{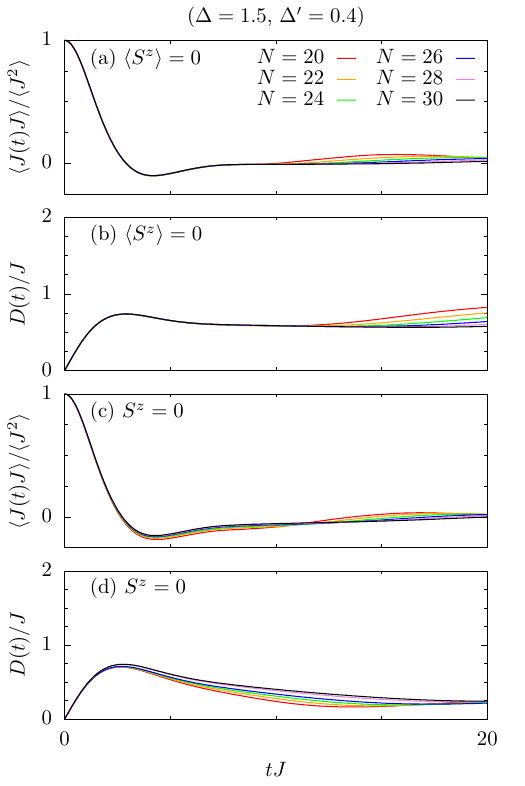}
\caption{The same data as in Fig.\ \ref{fig::finit_size_H} but now for
the model in Eq.\ (\ref{eq:model_NNN}) with $\Delta^{\prime} = 0.4$.}
\label{fig::finit_size_H_NNN}
\end{figure}

We now start the Lanczos iterative scheme, where we choose
as the seed operator $O_0$ the normalized current operator,
$O_0 = \tilde{j}$. First, we set
\begin{equation}
b_1 = \Vert {\cal L}O_0 \Vert \, , \quad |O_1) = \frac{{\cal L} |O_0)}{b_1} \, .
\end{equation}
Then, we iteratively compute
\begin{equation}
|O^\prime_n) = {\cal L} |O_{n-1}) - b_{n-1} |O_{n-2})
\end{equation}
and
\begin{equation}
b_n = || O^\prime_n || \, , \quad |O_n) = \frac{|O^\prime_n)}{b_n} \, .
\end{equation}
In this way, all Lanczos coefficients $b_n$ can in principle be calculated. In
practice, however, only the first few $b_n$ are easily numerically accessible.
To this end, we employ the operator growth hypothesis \cite{EhudPRX19}, which
states that, in a nonintegrable system, the $b_n$ for sufficiently large $n$
are asymptotically linear. Assuming \cite{smooth}
\begin{equation} \label{eq:linear_bn}
b_n = \alpha n + \beta \, , \quad n \ge R \, ,
\end{equation}
one obtains the following approximation of $F(0)$:
\begin{equation}
F_{0} \approx D_R = \begin{cases}
\frac{1}{\tilde{p}_{R}b_{R}}\prod_{m=1}^{\frac{R}{2}}\frac{b_{2m}^{2}}{b_{2m-1}^
{2}} & \text{, even }R\\
\frac{\tilde{p}_{R}}{b_{R}}\prod_{m=1}^{\frac{R-1}{2}}\frac{b_{2m}^{2}}{b_{2m-1}
^{2}} & \text{, odd }R
\end{cases}\, ,
\end{equation}
where
\begin{equation}
p_{R}=\frac{\Gamma(\frac{R}{2}+\frac{\beta}{2\alpha})\Gamma(\frac{R}{2}+\frac{
\beta}{2\alpha}+1)}{\Gamma^{2}(\frac{R}{2}+\frac{\beta}{2\alpha}+\frac{1}{2})}
\, .
\end{equation}
Here, we employ a simple approach, where $\alpha$ and $\beta$ are determined
by $b_R$ and $b_{R-1}$ only, i.e.,
\begin{equation}
\alpha_R = b_{R} - b_{R-1} \, , \quad \beta_R=Rb_{R-1}-(R-1)b_R \, .
\end{equation}

In Fig.\ \ref{fig::DR}, we summarize the so obtained $D_R$.
As  mentioned in the main text, $D_R$ is well converged w.r.t.\ $R$, for the
two models and the perturbation strengths considered.


\subsection{Dynamical quantum typicality}

Here, we briefly describe our typicality approach to the time dependence of
current autocorrelation functions in finite systems. More information on this
approach can be found in Refs.\ \cite{Heitmann2020, Jin2021}.

Loosely speaking, the concept of dynamical quantum typicality allows us to
replace expectation values w.r.t.\ ensembles by expectation
values w.r.t.\ single pure states, which are drawn at random from a
high-dimensional Hilbert space. In the context of current autocorrelation
functions, this concept leads to the approximation
\begin{equation} \label{eq:DQT}
\langle j(t) j \rangle = \frac{\langle \psi(t) | j | \varphi(t) \rangle}{\langle
\psi(0) | \psi(0) \rangle} + {\cal O} \left ( \frac{1}{\sqrt{d}} \right ) \, ,
\end{equation}
where $d = 2^N$ is the dimension of the Hilbert space. The two auxiliary pure
states read
\begin{equation}
\psi(t) = e^{- i H t} \, | \phi \rangle \, , \quad \varphi(t) = e^{- i H t}
j \, | \phi \rangle \, ,
\end{equation}
and $| \phi \rangle$ is a Haar-random pure state. Importantly, the time
argument in the approximation in Eq.\ (\ref{eq:DQT}) is now a property
of the two pure states and not of the current operator anymore. These pure
states can be propagated in time using fourth-order Runge-Kutta or Chebyshev
polynomials, which then gives access to comparatively large system sizes outside
the range of standard exact diagonalization. An approximation similar to the
one in Eq.\ (\ref{eq:DQT}) can be also obtained for other operators, which we
employ for the spatial-temporal correlation functions discussed later in the
context of open systems.

\begin{figure}[t]
\includegraphics[width=0.8\columnwidth]{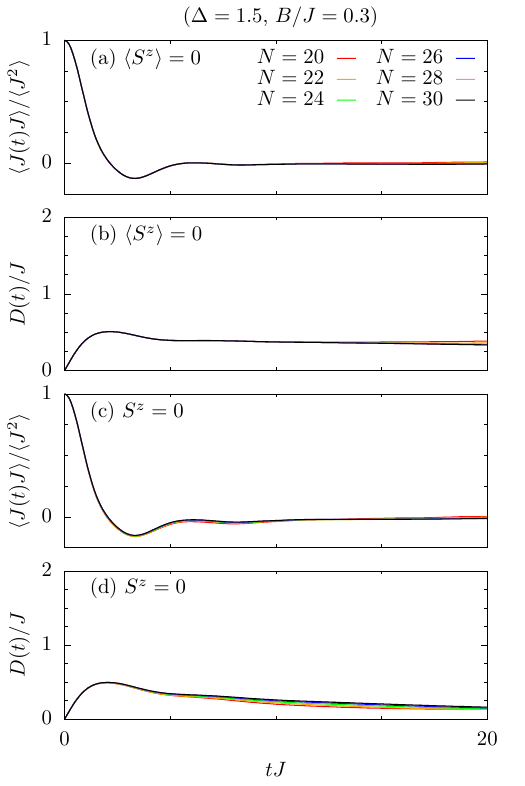}
\caption{The same data as in Fig.\ \ref{fig::finit_size_H} but now
for the model in Eq.\ \eqref{eq:model_B} with $B/J = 0.3$.}
\label{fig::finit_size_H_stagg}
\end{figure}

In Figs.\ \ref{fig::finit_size_H} - \ref{fig::finit_size_H_stagg}, we depict
raw data for the current autocorrelation function $\langle j(t) j \rangle$ and
its time integral $D(t)$ for three examples: The integrable model in Eq.\
(\ref{eq:model}) and the nonintegrable models in Eqs.\ (\ref{eq:model_NNN})
[with $\Delta' = 0.4$] and (\ref{eq:model_B}) [with $B/J = 0.3$]. We do so for
the two ensembles, grandcanonical $\langle S^z \rangle = 0$ and canonical
$S^z = 0$. This and similar raw data is the basis of our analysis of
diffusion constants in the main text. It is worth mentioning that the
finite-size effects are unrelated to the approximation error in Eq.\
(\ref{eq:DQT}), which is negligibly small in all cases considered.

\begin{figure}[t]
\includegraphics[width=0.8\columnwidth]{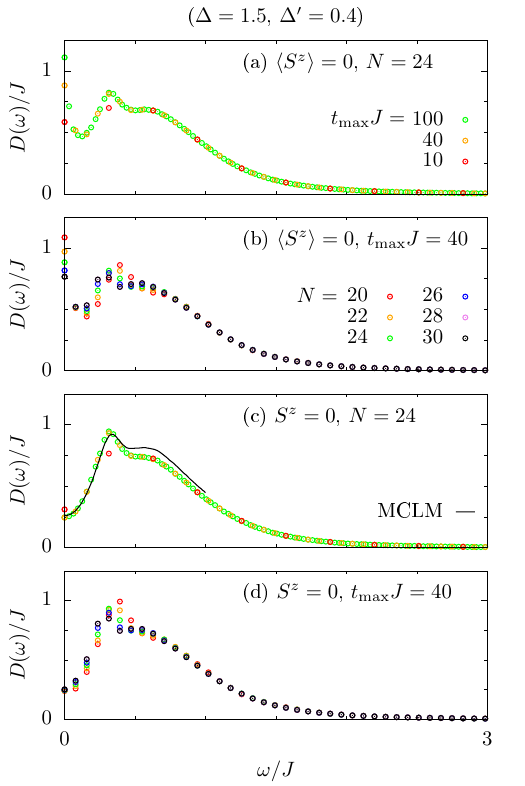}
\caption{Diffusion coefficient $D$ versus frequency $\omega$ for the model
in Eq.\ (\ref{eq:model_NNN}) with $\Delta^{\prime} = 0.4$, as obtained
numerically for the grandcanonical ensemble $\langle S^z \rangle = 0$ using (a)
fixed system size $N = 24$ and different times $t_{\text{max}}$ and (b) fixed
$t_{\text{max}}$ and different $N$. (c) and (d) show (a) and (b) for the
canonical ensemble $S^z = 0$. In (c), the MCLM result from Ref.\
\cite{Prelovsek2022} is also depicted.}
\label{fig::frequency_H_NNN}
\end{figure}

\begin{figure}[t]
\includegraphics[width=0.8\columnwidth]{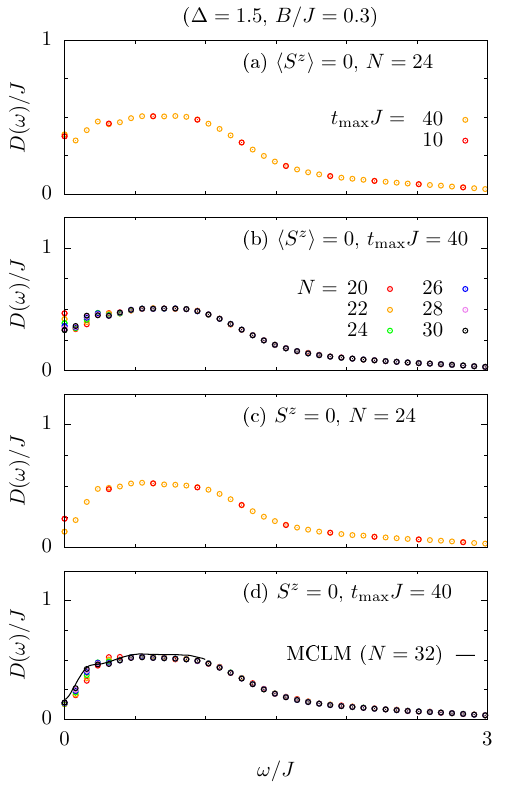}
\caption{The same data as in Fig.\ \ref{fig::frequency_H_NNN} but now for the
model in Eq.\ (\ref{eq:model_B}) with $B/J = 0.3$.}
\label{fig::frequency_H_stagg}
\end{figure}

In Figs.\ \ref{fig::frequency_H_NNN} and \ref{fig::frequency_H_stagg}, we
additionally depict for the two nonintegrable cases the current
autocorrelation function in frequency space, as obtained from a Fourier
transform up to a maximum time $T_\text{max}$. As can be seen, the specific
choice of $T_\text{max}$ controls the frequency resolution, and a reasonable
choice in a finite system is not obvious. As an additional consistency check of
our approach, we compare to existing data from the microcanonical Lanzos method
\cite{Prelovsek2022} and find a convincing agreement, as visible in Figs.\
\ref{fig::frequency_H_NNN} (c) and \ref{fig::frequency_H_stagg} (d).
MCLM data at zero frequency is also shown in the main text.


\section{Calculation of nonequilibrium transport}\label{ap::B}
\subsection{Time evolution of densities}

Here, we summarize the main aspects of our approach to the open-system
dynamics, which is a prediction just on the basis of the closed-system
dynamics. More details on this approach can be found in Refs.\
\cite{Heitmann2023a, Heitmann2023, Kraft2024}.

Specifically, the prediction is based on spatio-temporal correlation
functions at infinite temperature,
\begin{equation} \label{eq:spatio_temporal}
\langle S^{z}_{r}(t)S_{r'}^{z}(0)\rangle = \frac{\text{tr}[e^{i
H t}S^{z}_{r}e^{- i H t} S_{r'}^{z}]}{2^{N}} \; .
\end{equation}
Before we come to the actual prediction, it is useful to define the quantity
\begin{equation}
C_{r}(t) = \langle S^{z}_{r}(t)S_{1}^{z}(0)\rangle
- \langle S^{z}_{r}(t)S_{N/2 + 1}^{z}(0)\rangle \, ,
\end{equation}
which is the difference of two correlation functions with $r' = 1$ and $r' = N/2
+ 1$. Further, it is useful to introduce the superposition
\begin{equation} \label{eq::superposition}
d_{r}(t) = 2 \mu \sum_{j} A_{j} \, \Theta(t-\tau_{j}) \, C_{r}(t-\tau_{j})
\, ,
\end{equation}
where $A_{j}$ are some amplitudes, $\tau_{j}$ are some times, and $\Theta(t)$
is the Heavyside function. Using this notation, the prediction for the
open-system dynamics can be written as \cite{Heitmann2023}
\begin{equation}
\langle S_{r}^{z}(t)\rangle \approx \frac{1}{T_{\text{max}}}
\sum_{T=1}^{T_{\text{max}}} d_{r,T}(t) \, ,
\label{eq::prediction}
\end{equation}
where the sum runs over $T_{\text{max}}$ different time sequences
$(\tau_{1},\tau_{2},\dots)$. Here, a particular time sequence is
generated via
\begin{equation}
\tau_{j+1} = \tau_{j} - \frac{\ln(\varepsilon_{j+1})}{2\gamma} \, ,
\end{equation}
where $\varepsilon_{j+1}$ are random numbers drawn from a uniform distribution
$]0,1]$. The amplitudes  $A_{j}$ in Eq.\ \eqref{eq::superposition} read
\begin{equation}
A_{j} = \frac{a_{j}\ -\ d_{1,T}(\tau_{j} - 0^{+})}{\mu}
\end{equation}
with
\begin{equation}
a_{j} = \frac{\mu\ -\ 2\ d_{1,T}(\tau_{j} - 0^{+})}{2\ -\ 4\mu\
d_{1,T}(\tau_{j} - 0^{+})} \; .
\end{equation}
The accuracy of this prediction has been demonstrated for
integrable \cite{Heitmann2023}
and nonintegrable \cite{Kraft2024} systems, and it is particularly high for
periodic boundary conditions.

\subsection{Current in the steady state}

\begin{figure}[t]
\includegraphics[width=0.9\columnwidth]{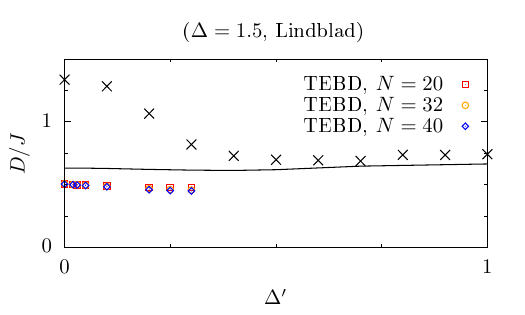}
\caption{Open boundary conditions and large system-bath coupling. TEBD
data is compared to the results in Fig.\ \ref{fig:diffusion_open} (b) of the
main text.} \label{fig:TEBD}
\end{figure}

To obtain the current in the steady state, we need the injected magnetization
$\langle \delta S_{1}^{z}\rangle$ at the first bath site, which can be
predicted as \cite{Heitmann2023}
\begin{eqnarray}
\langle \delta S_{1}^{z}(t)\rangle \approx \frac{1}{T_{\text{max}}}
\sum_{T=1}^{T_{\text{max}}} \delta d_{1,T}(t)
\end{eqnarray}
with
\begin{eqnarray}
\delta d_{1,T}(t) = 2 \mu \sum_{j} A_{j} \Theta(t-\tau_{j})\
\langle [S_{1}^{z}(0)]^{2}\rangle
\; .
\end{eqnarray}
In the steady state, all local currents are the same,
\begin{eqnarray}
\langle j_{r}\rangle = \langle j_{r'}\rangle \, , \quad 1 \leq r,r' \leq
N/2 +1 \, .
\end{eqnarray}
Thus, it is sufficient to know $\langle j_{1}\rangle$ to get all other
local currents. In particular, $\langle j_{1} \rangle$ can
be obtained from the injected
magnetization via \cite{Heitmann2023}
\begin{eqnarray}
\langle j_{1}\rangle = \frac{\text{d}}{\text{d} t} \frac{\langle
\delta S_{1}^{z}(t)\rangle}{2}\; ,
\end{eqnarray}
where the factor $1/2$ takes into account that the injected magnetization can
flow to the left and to the right of this bath, due to periodic boundary
conditions. Then, we can compute the diffusion constant in the
steady state by
\begin{eqnarray}
D = - \frac{\langle j_{1} \rangle}{\langle S_{r+1}^{z}\rangle -
\langle S_{r}^{z}\rangle}\; ,
\end{eqnarray}
for some $r$ in the bulk of the system.

\subsection{Open boundary conditions}

While we have focused throughout the main text on the case of periodic boundary
conditions, it is certainly an interesting and important question whether and
to what degree the results change for the case of open boundary conditions.
Unfortunately, our approach to open systems is less reliable in this case.
Instead, we show in Fig.\ \ref{fig:TEBD} results from time-evolving block
decimation (TEBD) for a large system-bath coupling $\gamma/J = 1$ and compare to
the results in Fig.\ \ref{fig:diffusion_open} (b) of the main text.

Consistent with the simple model, the TEBD data do not show an increased value
of $D(\Delta')$ for small $\Delta'$, as there is no finite-size Drude weight
for open boundary conditions. Notably, $D(\Delta')$ at fixed $\Delta$
decreases, as the system size $N$ is increased, at least for the system sizes
depicted. While it is hard to conclude on the behavior in the thermodynamic
limit, the comparison indicates a sensitivity on the specific boundary
conditions employed, which might be traced back to an anomalous character of
the integrable point.

\begin{figure}[t]
\includegraphics[width=0.9\columnwidth]{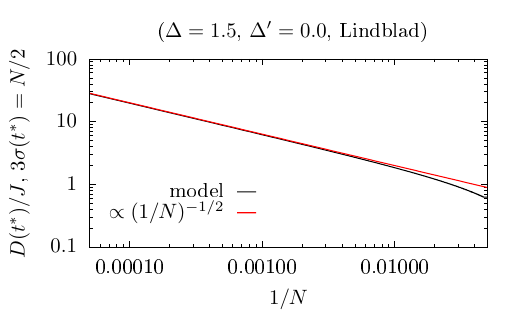}
\caption{The same data as in Fig.\ \ref{fig:model}, but now depicted in a
log-log plot and compared to the power law $\propto (1/N)^{-1/2}$.}
\label{fig:asymptotic}
\end{figure}

\section{Details on the simple model} \label{ap::C}

\subsection{Time-dependent diffusion coefficient}

In the main text, we have briefly introduced a simple model for the system-size
scaling of the diffusion constant in the open system. Thus, we provide
here an extended discussion and further details.

The simple model is based on the closed system and makes an assumption on the
time-dependent diffusion coefficient in Eq.\ (\ref{eq:D_LRT}),
\begin{equation} \label{eq:D_LRT_repeat}
D(t) = \frac{1}{\chi} \int_{0}^{t} \text{d}t' \, \langle j(t') j \rangle \, .
\end{equation}
Specifically, this assumption reads
\begin{equation} \label{eq:assumption}
D(t) = a t \, \Theta(t_1 - t) + b \, \Theta(t - t_1) + c \, (t - t_2) \,
\Theta(t - t_2) \, ,
\end{equation}
where $\Theta(t)$ is the Heavyside function. Hence, $D(t)$ first increases
linearly up to a time $t_1$, then remains constant up to a time $t_2$, and
finally increases linearly again. In particular, as illustrated in Fig.\
\ref{fig:model}, the quantities $t_2$ and $c$ are assumed to depend on system
size as \cite{Steinigeweg2012}
\begin{equation}
t_2 \propto N \, , \quad c \propto \frac{1}{N} \, ,
\end{equation}
while the other quantities $t_1$, $a$, and $b$ are independent of system size.
The parameter $a$ is given by
\begin{equation}
a = \frac{\langle j^2 \rangle}{\chi} = \frac{1}{2} \,
\end{equation}
and the remaining parameters are obtained from a fit to $N = 20$ data for
$\Delta = 1.5$,
\begin{equation}
t_1  = 1.2 \, , \quad t_2 = 0.5 \, L \, , \quad  b = 0.63 \, , \quad c =
\frac{1.4}{N} \, .
\end{equation}
Note that the parameter $c$ results from a saturation of the current
autocorrelation at long times,
\begin{equation}
\lim_{t \to \infty} \frac{\langle j(t) j \rangle}{\chi} \propto \frac{1}{N} \, ,
\end{equation}
and plays the role of the finite-size Drude weight, which is
expected to vanish for $\Delta = 1.5$ and $N \to \infty$ \cite{Bertini2021}.
Thus, for $N \to \infty$, the assumption in Eq.\ (\ref{eq:assumption}) leads to
\begin{equation}
\lim_{t \to \infty} D(t) = b = \text{const.} \, ,
\end{equation}
which indicates normal diffusion, as expected for $\Delta = 1.5$
\cite{Bertini2021}. It is worth pointing out that the finite-size Drude weight
could in principle hide anomalous behavior,
\begin{equation}
\frac{\langle j(t) j \rangle}{\chi} \propto \frac{1}{t} \, ,
\end{equation}
and therefore a logarithmic divergence of the diffusion coefficient,
\begin{equation}
D(t) \propto \log t \, .
\end{equation}
However, numerical simulations like ours in Fig.\ \ref{fig::finit_size_H} do
not support such an anomalous contribution and already indicate a constant
plateau. Moreover, for the following discussion, a potentially anomalous
contribution plays a minor role. In fact, the finite-size Drude weight is much
more important.

\subsection{Real-time broadening}

Next, let us stay in the closed system and consider the spatio-temporal
correlation functions $\langle S^{z}_{r}(t)S_{r'}^{z}(0)\rangle$ in Eq.\
(\ref{eq:spatio_temporal}). At $t = 0$,
\begin{equation}
\langle S^{z}_{r}(0)S_{r'}^{z}(0)\rangle = \frac{1}{4} \, \delta_{r,r'}
\end{equation}
at infinite temperature, where $\delta_{r,r'}$ denotes the Kronecker symbol.
Therefore, the profile is initially a sharp peak at the lattice site $r = r'$.
The real-time broadening of this peak can be followed via the spatial variance,
\begin{eqnarray}
\sigma^2(t) \! &=& \! \sum_r (r-r')^2 \, \frac{\langle
S^{z}_{r}(t)S_{r'}^{z}(0)\rangle}{1/4} \nonumber \\
&+& \! \Big [\sum_r (r-r') \, \frac{\langle
S^{z}_{r}(t)S_{r'}^{z}(0)\rangle}{1/4} \Big ]^2 \, ,
\end{eqnarray}
which can be written in the form of Eq.\ (\ref{eq:variance}),
\begin{equation} \label{eq:variance_repeat}
\sigma^2(t) = 2 \int_0^t \text{d}t' \, D(t') \, ,
\end{equation}
and is expressed in terms of the time-dependent diffusion coefficient $D(t)$ in
Eq.\ (\ref{eq:D_LRT}). Note that this expression does not require the
assumption for $D(t)$ in Eq.\ (\ref{eq:assumption}). Now, let us consider the
time $t^*$, when the intially sharp peak has broadened over half of the system,
\begin{equation}
3 \sigma(t^*) = \frac{N}{2} \, ,
\end{equation}
where we take three times the standard deviation for sake of simplicity. Using
the assumption in Eq.\ (\ref{eq:assumption}), $t^*$ can then be calculated for
any system size $N$. For large $N$, one finds the asymptotic behavior
\begin{equation}
t^* \propto N^{3/2}
\end{equation}
and, in particular, $t^* \gg t_2$. For normal diffusion, as a side
remark, one would have $t^* \propto N^2$.

\subsection{Link to the open system}

Eventually, let us link to the actual open system. To
this end, we assume that,
for the open system, (i) $t^*$ is the relevant time scale and that (ii)
$D(t = t^*)$, evaluated at $t = t^*$, is the corresponding diffusion
coefficient. Note that these two assumptions are reasonable, in view of the
correspondence between closed and open systems in and around Eq.\
(\ref{eq::prediction}).

In Fig.\ \ref{fig:model}, we have shown the finite-size scaling of the so
obtained $D(t^*)$. On the one hand, $D(t^*)$ agrees nicely with the $1/N$
scaling, as suggested by numerics for small system sizes, which serves as a
consistency check. On the other hand, $D(t^*)$ departs from a $1/N$
scaling for larger system sizes. To conclude on the asymptotic behavior, we
show in Fig.\ \ref{fig:asymptotic} the same data as in Fig.\ \ref{fig:model},
but now depicted in a log-log plot. Clearly,
\begin{equation}
D(t^*) \propto \sqrt{N} \, ,
\end{equation}
which can be also seen easily by plugging the asymptotic behavior of the time
$t^* \propto N^{3/2}$
into the assumption for the diffusion coefficient in Eq.\
(\ref{eq:assumption}).

We should stress that the assumption in Eq.\ (\ref{eq:assumption}) is only
applicable to the integrable point. For the case of a weak but nonvanishing
perturbation, this assumption is only expected to hold up to a certain system
size. Above this system size, the scaling of the Drude weight changes from a
power law to an exponential decrease in system size, which is consistent with
the eigenstate thermalization hypothesis \cite{Steinigeweg2013} and in turn
leads to a convergence of the diffusion coefficient w.r.t.\ system size.



\begin{thebibliography}{62}
\bibitem [{lin()}]{line}%
  \BibitemOpen
  \href@noop {} {}\bibinfo {note} {Even though the results from the recursion
method are indicated by a solid line in Fig. \ref{fig:diffusion_NNN}, these
results are also obtained for a discrete set of perturbation strengths (see
Appendix \ref{ap::A}).}\BibitemShut {Stop}%
\end{thebibliography}

%

\newpage

\end{document}